\newcommand{\vecr}{\mbox{\boldmath$r$}}
\newcommand{\vecv}{\mbox{\boldmath$v$}}
\newcommand{\vecp}{\mbox{\boldmath$p$}}
\newcommand{\vece}{\mbox{\boldmath$e$}}
\newcommand{\veca}{\mbox{\boldmath$a$}}
\newcommand{\vecu}{\mbox{\boldmath$u$}}

\newcommand{\vecF}{\mbox{\boldmath$F$}}

\newcommand{\dfd}{{\rm d}}

\documentclass{article}
\usepackage[]{graphicx}\usepackage[]{epsfig}
\begin{document}

\title{Gravitational Lagrangians, Mach's principle, and the equivalence principle in an expanding universe}
\author{Hanno Ess\'en
\\KTH Mechanics  \\ SE-100 44 Stockholm, Sweden\\
e-mail: hanno@mech.kth.se}
\date{December 30 2013}
\maketitle

\begin{abstract}
The gravitational Lagrangian based on special relativity and the assumption of a fourth rank tensor interaction, derived by Kennedy (1972), is used to check Mach's principle in a homogeneous isotropic expanding universe. The Lagrangian is found to be consistent with Mach's principle when the density is the critical density and inertial mass is suitably renormalized. The Kennedy approach only gives the Lagrangian to first order in the gravitational coupling constant. By invoking the equivalence principle higher order corrections are found which renormalize the gravi\-tational masses to the same values as the inertial masses. It is not the same as the correction derived from general relativity by Einstein-Infeld-Hoffmann, but otherwise the Lagrangians agree.
\end{abstract}

\section{Introduction}
Some connections between gravitational Lagrangian formalisms, cosmology, the critical density, Mach's principle, and the equivalence principle, are pointed out. It is first noted that, assuming the cosmological validity of a class of gravitational Lagrangians, a particle in an isotropic Hubble expanding universe will obey Mach's principle precisely when the density of the universe is the critical density ($\Omega = 1$). Mach's principle is defined so that the translational acceleration appearing in Newton's second law is the acceleration relative to that of the universe as a whole. It is pointed out that this is self consistent in the sense that the Lagrangian formalism assumes a background flat three-dimensional space, and this is consistent with general relativity precisely when the density is critical. The Lagrangian formalism gives this result for a renormalized inertial mass of the particle. In order for the equivalence principle to be valid it is then necessary that the masses occurring in the gravi\-tational interaction are renormalized in the same way. This is seen to require that the Lagrangian contains terms up to third order in the gravitational coupling constant (six-body interactions).

\section{Gravitational Lagrangians}
Fock \cite{BKfock} found the Lagrangian that yields the Einstein-Infeld-Hoffmann (EIH) equations of motion \cite{EIH,eddington&clark,BKinfeld&plebanski}. Modern derivations and discussions of this approach can be found in Landau and Lifshitz \cite{BKlandau2}, Hirondel \cite{hirondel}, Nordtvedt \cite{nordtvedt}, Brumberg \cite{brumberg}, and Louis-Martinez \cite{louis-martinez}, among others. These are all based on general relativity and Hirondel's is the shortest.

Here we will, however, focus on the profound work by Kennedy \cite{kennedy} on approximately relativistic interactions and their Lagrangians. Kennedy first derives the (special) relativistic Lagrangian for one particle interacting with another particle with constant given velocity. In a second step one then wishes to combine to such Lagrangians into a single two-body Lagrangian, symmetric in the particle indices. To do this it is necessary to expand the Lagrangians in $v/c$ and keep terms to second order only. Even then the desired Lagrangian can, in general, be found only to first order in the coupling constant, the exception being electromagnetism for which the formalism produces the Darwin Lagrangian \cite{darwin}.

Assuming that the interaction is mediated by a fourth rank tensor (presumably the curvature tensor) Kennedy arrives at a two body Lagrangian, which when generalized to $N$ bodies, is
\begin{equation}\label{eq.eih.lagr}
L = L_0 + L_1 + L_2 ,
\end{equation}
where,
\begin{equation}\label{eq.eih.lagr0}
L_0 = \sum_a \frac{ m_a \vecv_a^2}{2}  + G\sum_{a<b} \Phi(r_{ab}) ,
\end{equation}
and
\begin{equation}\label{eq.eih.lagr1}
L_1 = \sum_a \frac{m_a \vecv_a^4 }{8c^2}  +
\frac{G}{2c^2} \sum_{a<b} \ \Phi(r_{ab}) \left[ 3(\vecv_a^2+\vecv_b^2)-7\vecv_a\cdot\vecv_b - \frac{(\vecv_a\cdot\vecr_{ab}) (\vecv_b\cdot\vecr_{ab})}{r_{ab}^2} \right].
\end{equation}
Here $\vecr_{ab} =\vecr_b - \vecr_a$ is the vector from particle $a$ to particle $b$, and $r_{ab} = |\vecr_{ab}|$, and if $L_0$ is to give the Newtonian result we must have
\begin{equation}\label{eq.pot.minus.G}
\Phi(r_{ab}) = \frac{m_a m_b}{r_{ab}}.
\end{equation}
$L_2$ contains higher order terms in the gravitational coupling constant $G$,
\begin{equation}\label{e2.L2.unknown}
L_2 = G^2 L^* .
\end{equation}
To proceed we assume in what follows that $L_2$ does not depend on particle velocities. It will then not affect the inertia properties of a particle, {\it i.e.}\ the generalized momentum.

\section{Mach's principle}
Mach's principle has been subject of many publications over the years. Some more recent studies can be found in the volume edited by Barbour and Pfister \cite{BKbarbour&pfister}. Mashhoon et al.\ \cite{mashhoon&al} and Iorio et al.\ \cite{iorio&al} discuss the gravitomagnetic analogy. Other texts of interest are by Assis \cite{BKassis2}, Ciufolini and Wheeler \cite{BKciufolini&wheeler}, Peacock \cite{BKpeacock}, and by Cheng \cite{BKcheng_tp}. Frame dragging, rotational and translational, and its relation to Mach's principle and general relativity has been discussed by Gr{\o}n \cite{gron,hartman}, Gr{\o}n and Eriksen \cite{gron&eriksen}, Harris \cite{harris}, Holstein \cite{holstein}, Hughes \cite{hughes}, Lynden-Bell et al. \cite{lyndenbell&bicak&katz}, Mart\'{i}n et al.\ \cite{martin&al}, Nightingale \cite{nightingale,nightingale&ray}, and by Vet\H{o} \cite{veto1,veto2}. Recent experimental support was found by Everitt et al.\ \cite{everitt&al}.

Here we will interpret Mach's principle as saying that the equation of motion of a (slow) particle number $1$ is of the form,
\begin{equation}\label{eq.mach.princip.precise}
m_1 (\veca_1 - \dot{\vecu}) = \vecF_1 ,
\end{equation}
where $\dot{\vecu}$ is the acceleration of the universe as a whole. {\it I.e.}\ it is in the Newtonian form but only the acceleration relative to the universe as a whole, $\veca_1 - \dot{\vecu}$, is what matters in the equation of motion. The force $\vecF_1$ is from the other $N-1$ particles of the system. We will now show that this result can be obtained from the Lagrangian of Eq.\ (\ref{eq.eih.lagr}).

The equation of motion for particle $1$ is given by,
\begin{equation}\label{eq.euler.lagrange}
\frac{\dfd}{\dfd t} \frac{\partial L}{\partial \vecv_1} = \frac{\partial L}{\partial \vecr_1} \; \Leftrightarrow\; \dot{\vecp}_1 = \vecF_1 .
\end{equation}
All terms involving accelerations will occur on the left hand side here, so this is what we need to calculate. Calculation gives (assuming $L_2$ does not depend on velocities),
\begin{equation}\label{eq.gen.momentum}
\vecp_1 \equiv \frac{\partial L}{\partial \vecv_1} = m_1 \vecv_1 + \frac{G m_1}{2 c^2} \sum_{b=2}^N \frac{m_b}{r_{1b}} \left[ 6\vecv_1 - 7\vecv_b - \frac{\vecr_{1b} (\vecv_b \cdot \vecr_{1b}) }{r_{1b}^2}\right],
\end{equation}
for the generalized momentum $\vecp_1$. We now investigate under what circumstances one finds that $\dot{\vecp}_1 = m_1 (\veca_1 -\dot{\vecu})$, {\it i.e.} Mach's principle as stated in Eq.\ (\ref{eq.mach.princip.precise})?

\begin{figure}[h] \centering\includegraphics[width=200pt]{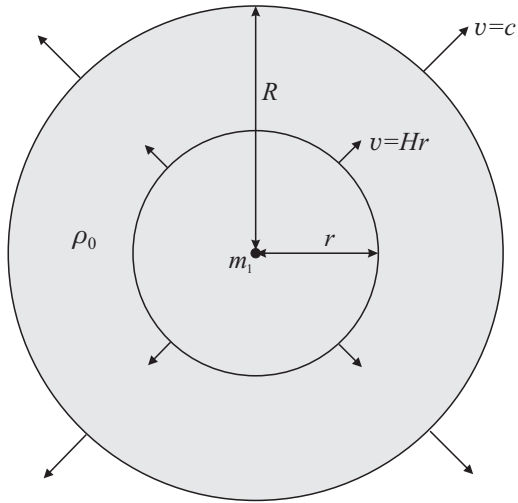}\vspace{1ex}
\caption{This figure illustrates the central point particle of mass $m_1$ in a homogeneous isotropic Hubble expanding universe with constant rest mass/energy density $\rho_0$. The radial speed is $v(r)=Hr$ and at the Hubble radius $R$ the speed is $v(R)=c$, the speed of light. In the Lagrangian we assume that the particle has velocity $\vecv_1$ and that the expanding sphere of density $\rho_0$ has overall translational velocity $\vecu$.} \label{FIG1}
\end{figure}

\section{Summing over an expanding universe} \label{sec.integr}
Now assume that particle 1 is in a homogeneous isotropic expanding universe of (constant rest mass/energy) density $\rho_0$, and with Hubble parameter $H$. The particles $m_b$ are then replaced by mass elements $\rho(r)\, \dfd V$ of position $\vecr$ and velocity $\vecv = H \vecr + \vecu$. Here $\vecu$ is an overall velocity of the universe relative to the origin, and we assume that it is small compared to the speed of light ($u\ll c$). We can then replace the sum in (\ref{eq.gen.momentum}) with an integral and get
\begin{equation}\label{eq.gen.momentum.int}
\vecp_1 = m_1 \vecv_1 + \frac{G m_1}{2 c^2} \int \frac{\rho(r)}{r} \left[ 6\vecv_1 - 7(H \vecr + \vecu) - \frac{\vecr (H r^2 + \vecu\cdot\vecr) }{r^2}\right]\, \dfd V.
\end{equation}
Clearly one can not assume that $Hr \ll c$ for all $r$, but because of the isotropy these velocities will not appear in the Lagrangian after the integration, and we will only be interested in the effective Lagrangian for slow particles in the expanding universe.

We now calculate the integral on the right hand side. Introduce spherical coordinates $(r, \varphi, \theta)$ and do the integration over the visible universe. At the radius $R$ of the visible universe the Hubble expansion leads to recession at the speed of light, $HR=c$, see Fig.\ \ref{FIG1}. The volume element in spherical coordinates is $\dfd V = r^2 \sin\theta\, \dfd r\, \dfd\varphi\, \dfd\theta$. Without loss of generality we assume that $\vecu = u\vece_z$. Since $\vecr = r (\sin\theta\,\vece_{\varphi} + \cos\theta\,\vece_z)$, where $\vece_{\varphi}=\cos\varphi\,\vece_x + \sin\varphi\,\vece_y$, the scalar product term becomes
\begin{equation}\label{eq.calc.sclar}
\vecr (\vecu\cdot\vecr) = r^2 u (\sin\theta\,\vece_{\varphi} + \cos\theta\,\vece_z) \cos\theta.
\end{equation}
The integrations over the sphere of radius $R$ will make the terms involving $H$ vanish for symmetry reasons, since these are multiplied by $\vecr$. Nothing depends on the angle $\varphi$ in the integral so the term multiplying $\vece_{\varphi}$ also vanishes. Two different integrals then remain to calculate, first
\begin{equation}\label{eq.pot}
\int \frac{\rho\, \dfd V}{r} = 4\pi  \int_0^R \rho(r) r \dfd r  ,
\end{equation}
and then,
\begin{equation}\label{eq.int.scal}
\int \frac{\rho \cos^2\theta\, \dfd V}{r} = 2\pi \int_0^R \rho(r) r\, \dfd r \int_0^{\pi} \cos^2\theta\, \sin\theta \,\dfd\theta = \frac{4\pi}{3} \int_0^R \rho(r) r \dfd r ,
\end{equation}
due to the scalar product term. Here the $\theta$-integral is $2/3$. In a previous study (Ess\'en \cite{essen13}) the mass/energy density $\rho(r)$ in the integral
\begin{equation}\label{eq.rad.int}
I[\rho]=\int_0^R \rho(r) r\, \dfd r
\end{equation}
was taken constant, $\rho(r) =\rho_0$, with the result that the integral evaluated to $I=\rho_0 R^2/2$. It is, however, physically more natural to assume that it is the {\em rest} mass/energy density that is constant. The expansion with radial speed $v(r)=Hr$ then means that one should use the special relativistic factor $1/\sqrt{1-v^2(r)/c^2}$ to get the contributing mass/energy from a shell of radius $r$, {\it i.e.}\ we put $\rho(r)=\rho_0 / \sqrt{1-v^2(r)/c^2}$. This gives,
\begin{equation}\label{eq.rad.int.rel}
I[\rho]=\int_0^R \frac{\rho_0 }{\sqrt{1-\frac{r^2}{R^2}}}\,r\, \dfd r = \rho_0 R^2 .
\end{equation}
The result of the radial integration is a then factor of 2 larger than was found previously \cite{essen13}, when the mass/energy density was simply assumed constant. One notes that Gogberashvili \cite{gogberashvili} has obtained this factor of two in a different but seemingly more arbitrary way.

\section{Generalized momentum density dependence}
Using the above results Eq.\ (\ref{eq.gen.momentum.int}) gives us
\begin{equation}\label{eq.gen.momentum.calculated}
\vecp_1 = m_1 \left[ \left(1+ 12 \frac{G\pi\rho_0 R^2}{c^2} \right)\vecv_1 - \left( \frac{44}{3} \frac{G\pi\rho_0 R^2}{c^2} \right) \vecu \right].
\end{equation}
Noting that the quantity,
\begin{equation}\label{eq.quatity.in.p}
\frac{G\pi\rho_0 R^2}{c^2} = \frac{G \pi}{H^2}\rho_0 ,
\end{equation}
using $R=c/H$, can be expressed as
\begin{equation}\label{eq.sigma.Omega}
\frac{G\pi\rho_0 R^2}{c^2} = \frac{3}{8} \frac{\rho_0}{\rho_c} \equiv \frac{3}{8} \Omega ,
\end{equation}
where
\begin{equation}\label{eq.rho.crit}
\rho_c = \frac{3H^2}{8\pi G}.
\end{equation}
is the critical density and where $\Omega$ is standard notation in cosmology for the ratio of the density to the critical density. In cosmology using general relativity and the assumption of an expanding homogeneous, isotropic universe one finds that the mass/energy density $\rho_c$ is the one that makes (three dimensional) space flat \cite{BKcheng_tp,jordan}. $\rho_c$  corresponds to the mass $M$ of the universe inside the Hubble radius $R=c/H$ being such that the Hubble radius is equal to the Schwarzschild radius $R=2GM/c^2$.

The generalized momentum (\ref{eq.gen.momentum.calculated}) of particle 1 now becomes,
\begin{equation}\label{eq.gen.mom.omega}
 \vecp_1 = m_1 \left[ \left(1+ \frac{9}{2} \Omega \right)\vecv_1 - \left( \frac{11}{2} \Omega \right) \vecu \right].
\end{equation}
Returning to my formulation of Mach's principle in (\ref{eq.mach.princip.precise}) it is seen to be realized with this $\dot{\vecp}_1$ if $\Omega = 1$. For this value of the density ratio the generalized momentum is
\begin{equation}\label{eq.gen.mom.omega.2}
 \vecp_1 = m_1 \frac{11}{2} (\vecv_1 - \vecu ) \; \Rightarrow \; \dot{\vecp}_1 = m_1 \frac{11}{2} (\veca_1 - \dot{\vecu} ).
\end{equation}
So, for $\Omega=1$ the acceleration in Newton's second law is relative to the acceleration $\dot{\vecu}$ of the universe as a whole. One also notes that the "bare" mass $m_1$ has been "renormalized" to $m'_1=11 m_1/2$.

\section{The principle of equivalence}
Let us now consider two slow particles near each other but far from all other local masses in the universe. As above we can sum over the expanding universe to get the influence of all the distant bodies on the two particles of interest. The result will be that their inertial masses are renormalized as found above. Assuming that the density is the critical density only the accelerations relative to the rest of the universe as a whole is relevant.

If, however, we now invoke the principle of equivalence, {\it i.e.}\ the equality of the inertial and gravitational masses we run into a problem\footnote{It can be proved that inertial and gravitational mass must be proportional within classical mechanics \cite{chubykalo&vlaev}}. The effective Lagrangian for two slow particles near each other will be given by $L_0$ of Eq.\ (\ref{eq.eih.lagr0}) as modified by the cosmologically integrated $L_1$ of (\ref{eq.eih.lagr1}) and should thus be
\begin{equation}\label{eq.newL0}
L_0^{\rm eff} = \frac{ m'_1 (\vecv_1-\vecu)^2}{2}+\frac{ m'_2 (\vecv_2-\vecu)^2}{2}  + G\frac{m_1  m_2}{r_{12}}
\end{equation}
with $m'_a =11 m_a /2$. This will not give the correct local Newtonian equations of motion unless we reinterpret $G$, but if we do that we have to redo the cosmological integrations and we are running in circles.

To get away from this problem it is clear that somehow also the gravitational masses must be renormalized by the cosmological integrations. It is easy to see that one way of doing this is to change $\Phi$ of Eq.\ (\ref{eq.pot.minus.G}) to,
\begin{equation}\label{eq.pot.renorm}
\Phi'(r_{ab}) = \frac{  \left(m_a+\frac{3G}{c^2} \sum_{c\neq a} \Phi(r_{ac}) \right) \left(m_b+\frac{3G}{c^2} \sum_{d \neq b} \Phi(r_{bd}) \right)}{r_{ab}}.
\end{equation}
This clearly results in $\Phi'(r_{ab}) = (11/2)^2 \Phi(r_{ab})$ if the summations/integrations are done in the same way as in Sec.\ \ref{sec.integr}. If we really change $\Phi$ we will also change $L_1$ and we have not achieved anything. Instead the changes can be incorporated as higher order terms in $G$, with the same net result.

Such a correction is consistent with the Kennedy \cite{kennedy} approach since it leads to correction terms $L_2$ of Eq.\ (\ref{e2.L2.unknown}) proportional to $G^2$ and $G^3$. These are seen to be,
 \begin{equation}\label{eq.eih.lagr2.expl}
L_2 = \frac{G^2}{c^2} \sum_{a<b} \left\{ 6   \frac{m_a \left[\sum_{d\neq b} \Phi(r_{bd})\right]}{ r_{ab} }  + \frac{9G}{c^2} \frac{\left[ \sum_{c\neq a} \Phi(r_{ac}) \right] \left[ \sum_{d\neq b} \Phi(r_{bd})\right] }{r_{ab} } \right\}.
\end{equation}
When the entire universe is included such corrections are not small. With this choice of $L_2$ the entire formalism is consistent with both Mach's principle, the critical density, and the equivalence principle.

\section{Conclusions}
Summarizing we find that the Lagrangians $L_0$ and $L_1$ together, after integration over the expanding universe, give a generalized momentum for a particle that only depends on the velocity of the particle relative to the universe as a whole, when the density is the critical density. The inertial mass is, however, renormalized to be $11/2$ larger that the naked mass. The Newtonian gravitational forces arising from Lagrangian $L_0$ then no longer obey the principle of equivalence, unless one reinterprets $G$, but this would require reinterpreting $L_1$. Instead the introduction of $L_2$ of Eq.\ (\ref{eq.eih.lagr2.expl}) solves the problem. Together with $L_0$ and integrations over the expanding universe $L_2$ results in the renormalized gravitational masses that agree with the renormalized inertial masses. The principle of equivalence is thus saved (to Newtonian order).

The fact that only $\Omega = 1$ is consistent with Mach's principle in the above formalism is interesting but could, of course, be an accident due to canceling errors. Other authors have obtained similar results by completely different routes \cite{martin&al,veto2,gogberashvili}. We note that we have not used general relativity at all in the above considerations. Kennedy \cite{kennedy} derives his gravitational Lagrangian only assuming special relativity and a fourth rank tensor interaction. His Lagrangian, however, is the same as the EIH-Lagrangian except for terms of higher order in the coupling constant $G$. We note that general relativity gives a different such higher order term than the one obtained here using the equivalence principle.\\

\noindent{\bf\Large Acknowledgements}\\
I am grateful Prof.\ Vet\H{o} for insisting that $\Omega=1$ is the value consistent with Mach's principle.


\end{document}